\documentclass{article}
\usepackage{spconf,amsmath,graphicx,url,bm,multirow,cite,hyperref,enumitem}

\title{A modularized neural network with language-specific output layers for cross-lingual voice conversion}
\name{Yi Zhou, Xiaohai Tian, Emre Y\i lmaz, Rohan Kumar Das and Haizhou Li}
\address{National University of Singapore, Singapore\\
yi.zhou@u.nus.edu, \{eletia, emre, rohankd, haizhou.li\}@nus.edu.sg
}

\begin{document}
%
\maketitle
\begin{abstract}
This paper presents a cross-lingual voice conversion framework that adopts a modularized neural network.
The modularized neural network has a common input structure that is shared for both languages, and two separate output modules, one for each language.
The idea is motivated by the fact that phonetic systems of languages are similar because humans share a common vocal production system, but acoustic renderings, such as prosody and phonotactic, vary a lot from language to language.
The modularized neural network is trained to map Phonetic PosteriorGram (PPG) to acoustic features for multiple speakers. 
It is conditioned on a speaker i-vector to generate the desired target voice.
We validated the idea between English and Mandarin languages in objective and subjective tests.
In addition, mixed-lingual PPG derived from a unified English-Mandarin acoustic model is proposed to capture the linguistic information from both languages. 
It is found that our proposed modularized neural network significantly outperforms the baseline approaches 
in terms of speech quality and speaker individuality, and mixed-lingual PPG representation further improves the conversion performance.

\end{abstract}
\section{Introduction}
\label{sec:intro}

Voice conversion (VC) aims to modify the speech of one speaker (source) to sound like that of another speaker (target). 
Cross-lingual VC is a special case where the source and target speakers speak different languages.
It is the enabling technology for many real world applications such as speech-to-speech translation \cite{ramani2016multi}, foreign language training \cite{qian2011frame}, and movie dubbing, etc. 

The early studies of cross-lingual VC rely on parallel data recorded from bilingual speakers~\cite{abe1990cross,abe1991statistical}.
These approaches utilize parallel data of the source and target speakers in the target speaker's language during training, and then convert the source speaker's utterances in the other language to the target voice.
However, such bilingual source speakers are not always available in practice.
Various alignment methods have been proposed to find the closest matching segments between non-parallel speech data across languages, for example, unit selection~\cite{hunt1996unit,duxans2006voice,sundermann2006textcross,sundermann2006text}, iterative alignment~\cite{erro2007frame,uriz2009voice,erro2010inca,qian2011frame} methods, and the vocal tract length normalization (VTLN) based phone mapping approaches~\cite{sundermann2003vtln,desai2009framework,qian2011frame}.
Yet, the converted speech quality is degraded due to inaccurate alignments~\cite{erro2010inca,tian2018average}.
Besides, eigenvoice-based technique~\cite{charlier2009cross,ramani2016multi} is also developed by adapting a pre-trained eigenvoice Gaussian mixture model (EV-GMM) with a few utterances from the target speaker in a different language.
Automatic speech recognition (ASR) is also employed to provide the phonetic information.
In~\cite{xie2016klvc}, Kullback-Leibler divergence (KLD) is calculated to map the senones between two languages.
In~\cite{gan2018mandarin}, the source utterance is first translated to the target language, which is then used to synthesize the target utterance.
Recently, neural network approaches are widely studied. 
Deep generative models like variational auto-encoder (VAE)~\cite{hsu2017voice,saito2018non} and generative adversarial networks (GANs)~\cite{sismanadaptive,kaneko2019cyclegan} without requiring parallel data are found effective for VC, though in the same language.

Phonetic PosteriorGram (PPG) has been proposed for VC~\cite{sun2016phonetic, ccicsman2017sparse,liu2018voice,tian2019wavenet} and also successfully applied to cross-lingual conversions~\cite{sun2016personalized}.
PPG is a frame-level phonetic information representation obtained from the acoustic model in a speaker-independent ASR system.
The conversion model is trained to map PPGs to the output acoustic features~\cite{sun2016phonetic}.
To enhance the converted speech quality, 
a recent study~\cite{zhou2019cross} investigates the use of bilingual PPG, which is formed by stacking two monolingual PPGs in source and target languages, as a linguistic representation of both languages for cross-lingual VC. 
Additionally, an average modeling approach~\cite{tian2018average,gao2019sing} is also proposed to leverage the linguistic and acoustic information from various speakers in different languages~\cite{zhou2019cross, zhou2019embedding}. 

\begin{figure*}[t]
	\includegraphics[width=\textwidth]{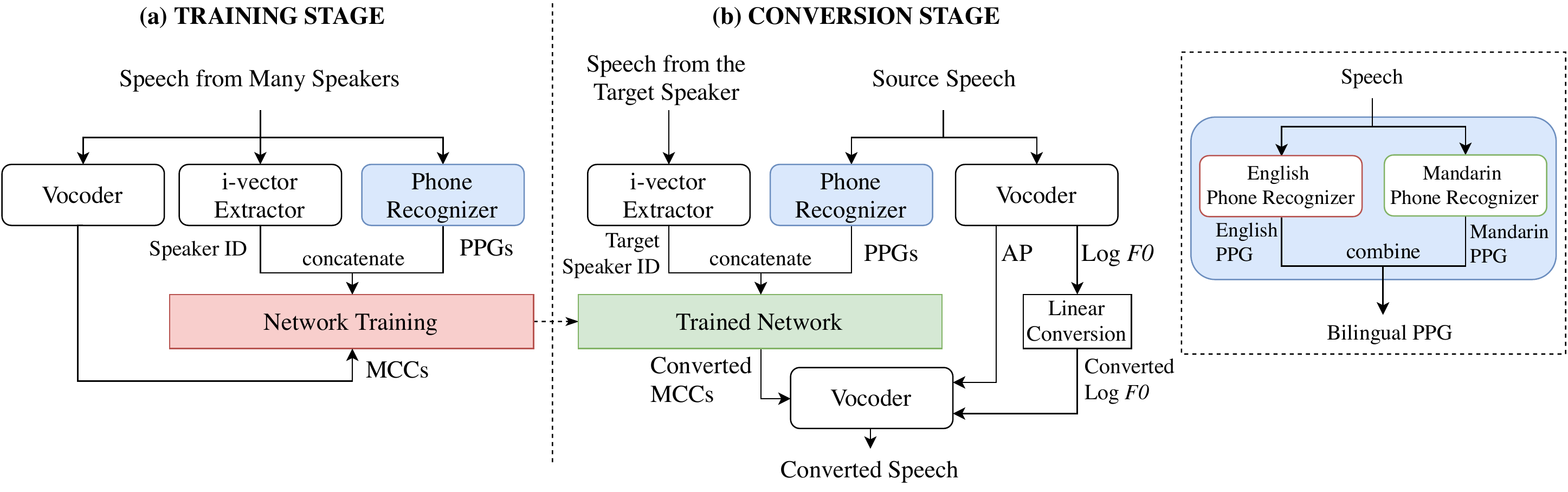}
	\vspace{-0.5cm}
	\caption{Block diagram of (a) training stage and (b) conversion stage of the PPG-based cross-lingual VC system. The English and
	Mandarin phone recognizers convert speech into PPGs that represent linguistic information, as illustrated in the dash box. Note that the PPGs are decoded by two phone recognizers separately, the combined PPG is called Bilingual PPG.}
	\label{fig:vc_system}
\end{figure*}

In this paper, we propose a modularized neural network using mixed-lingual PPG for cross-lingual VC.
Inspired by the language and speaker factorization acoustic modeling technique in multi-language and multi-speaker text-to-speech (TTS)~\cite{zen2012statistical,fan2016speaker,li2016multi},
we model the linguistic to acoustic feature transformation 
in two steps by a shared language-independent module, and two separate language-specific output modules.
The language-independent module is shared as a bridge to transfer input linguistic features from multiple speakers and languages into a common space.
While the separate language-specific modules are deployed to model the acoustic features for each language individually.
We hypothesize that the language specific output layers will improve the acoustic renderings of specific languages.
Mixed-lingual PPG is extracted by a jointly trained English-Mandarin mixed-lingual acoustic model, which is able to capture the acoustic information of both languages.
Hence, the assigned phone posterior probabilities are expected to reflect the shared characteristics of the two phonetic systems. 

\section{Related Work}
\label{sec:work}
This section presents the average modeling cross-lingual VC using bilingual PPG~\cite{zhou2019cross} and motivates our proposed work.

\subsection{Methodology}
\label{ssec:system}
Bilingual PPG is an effective characterization used to capture the phonetic information of two languages~\cite{zhou2019cross}.
Fig.~\ref{fig:vc_system} shows the block diagram of the bilingual PPG-based cross-lingual VC framework. 
To convert between English and Mandarin speakers, an English and a Mandarin phone recognizer are trained individually.
Illustrated in the dash box in Fig.~\ref{fig:vc_system}, monolingual PPGs are first extracted with English and Mandarin phone recognizer.
Then, the two monolingual PPGs are combined to form a bilingual PPG to represent the linguistic information of both languages.

As shown in Fig.~\ref{fig:vc_system}(a), during training, we first extract i-vector~\cite{dehak2010front}, linguistic features (PPGs), and acoustic features (mel cepstral coefficients (MCCs)) from bilingual speech data from multiple speakers.
Then, we form the input features by augmenting i-vector to PPGs.
The output features only contain MCCs. 
In this way, the mapping network $\mathcal{F}(\cdot)$ is trained to map input features ${\mathbf{X}=\{ \bm{x}_1, ...,\bm{x}_t,...,\bm{x}_T \} }$ to output features ${\mathbf{Y}=\{ \bm{y}_1, ...,\bm{y}_t,...,\bm{y}_T \} }$ by the back-propagation through time (BPTT) algorithm.
$T$ indicates the number of total frames in the training data.
The actual output values ${\mathbf{\hat{Y}}=\{ \bm{\hat{y}}_1, ...,\bm{\hat{y}}_t,...,\bm{\hat{y}}_T \} }$ are obtained according to the mapping network,
\begin{equation} \label{eq:1}
\mathbf{\hat{Y}} = \mathcal{F}(\mathbf{X})
\end{equation}

The conversion stage is shown in Fig.~\ref{fig:vc_system}(b). 
We first extract the PPGs and target i-vector from the source speech and the target speech using the same phone recognizer and i-vector extractor, respectively.
PPGs and i-vector are then concatenated to form the input features for the mapping network to convert the MCCs.
Finally, the converted speech can be synthesized with linearly converted \textit{F0}, source aperiodicity (AP) and converted MCCs.

\subsection{Motivations}
\label{ssec:limitation}
The successful implementation of the cross-lingual VC framework with bilingual PPG and average modeling motivates us to consider in two directions.
First, despite the fact that languages share the common vocal production system~\cite{ning2017learning, zen2012statistical}, their acoustic renderings differ very much from one to another, such as their prosodic and phonotactic renderings~\cite{li2013spoken}. 
As a result, using only a single output layer may not be able to encode the unique language details.
Second, bilingual PPG is obtained by combining English PPG and Mandarin PPG, which are extracted by two individually trained acoustic models.
As the acoustic models are trained separately, the resultant bilingual PPG may not provide a unified view on the input speech from two languages.
Moreover, it could be biased towards one of the languages. 

\section{Modularized Neural Network and mixed-lingual PPG}
\label{sec:proposed_techniques}
In this section, we propose a modularized neural network with language-specific output layers to model a two-step acoustic feature generation process.
Instead of using bilingual PPG, we introduce a mixed-lingual PPG estimated by a unified English-Mandarin acoustic model.
%
\vspace{-0.2cm}
\subsection{Modularized Neural Network}
\label{ssec:network}
\begin{figure}[t]
	\centerline{\includegraphics[width=0.5\textwidth]{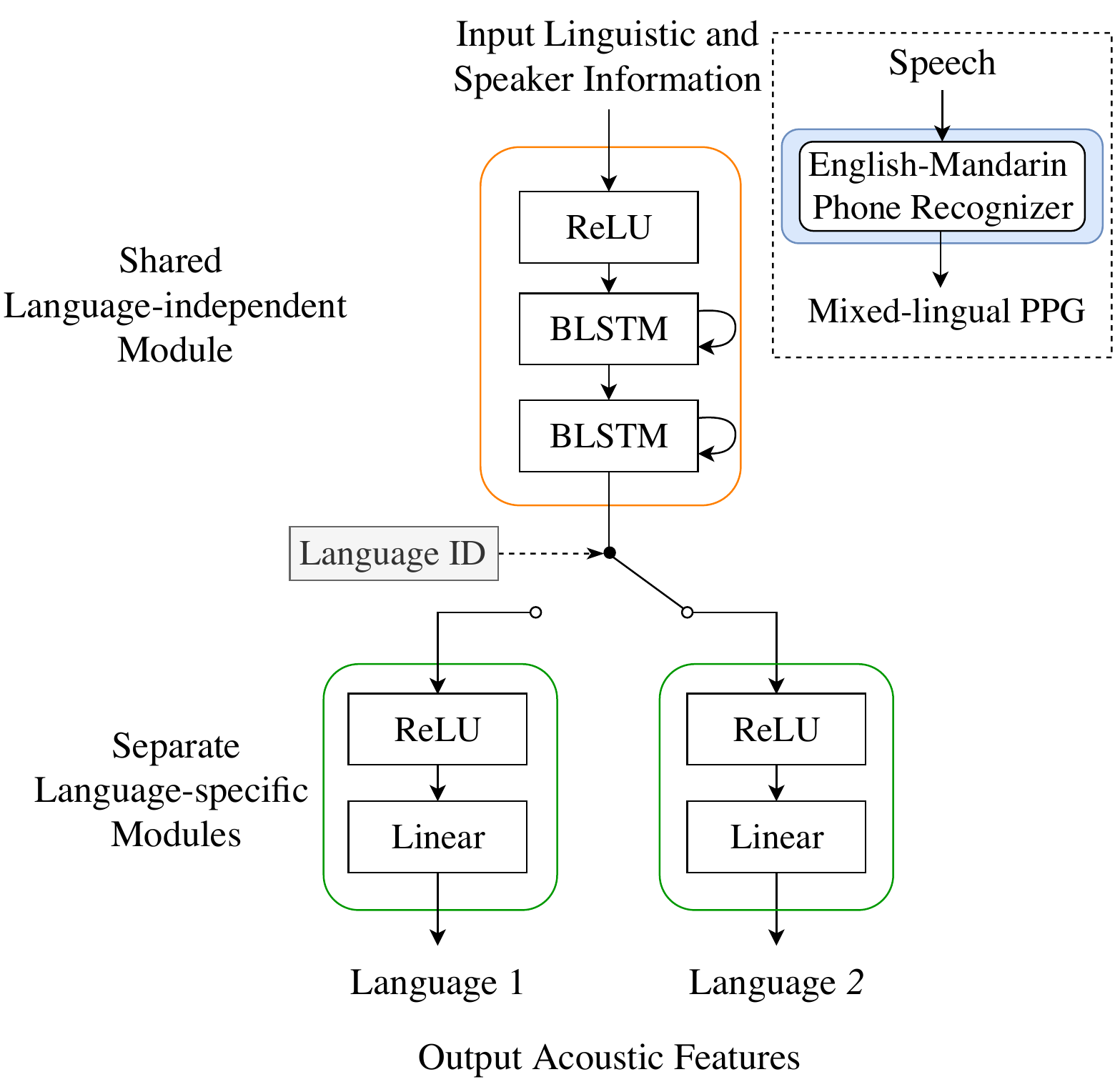}}
	\vspace{-0.4cm}
	\caption{The architecture of the proposed modularized neural network that employs multi-task learning.
	language-specific output layers. It consists of a shared language-independent module and two separate language-specific modules. Language ID serves as a switch to select the output module. The English-Mandarin phone recognizer converts speech into mixed-lingual PPGs that represent linguistic information, as illustrated in the dash box.}
	\label{fig:network}
\end{figure}
Multi-task learning is a learning paradigm in machine learning, which can effectively improve the generalization performance on all tasks by sharing representations between multiple related tasks~\cite{caruana1997multitask,zhang2017survey}.
Motivated by its success in many applications~\cite{ruder2017overview,zhang2019joint}, we propose a modularized neural network, as shown in Fig.~\ref{fig:network}. 
There are two functional blocks that represent the two steps in speech generation, namely shared language-independent module, and separate language-specific modules. 

In the shared language-independent module, both linguistic information (PPGs) and speaker information (i-vector) first pass through shared layers including a rectified linear activation function (ReLU) projection layer and two bidirectional long short-term memory (BLSTM) layers.
This shared module can capture the general transformation properties for both languages benefiting from multiple speakers' large database.

In the separate language-specific modules, each module consists of two language-specific output layers including a ReLU layer and a linear layer.
Given a language ID, language-independent intermediate features will be first forwarded to its corresponding ReLU layer to be mapped into language-dependent vectors.
The subsequent linear layer in the chosen language will then convert the learned complex features to the acoustic features.
The mapping network in Equation~(\ref{eq:1}) now can be expressed as 
\begin{equation}\label{eq:2}
  \mathbf{\hat{Y}} = \mathcal{F}(\mathbf{X}) =
    \begin{cases}
      {\mathcal{L}_{en} (\mathcal{S}(\mathbf{X}))}, & \text{English}\\
      {\mathcal{L}_{cn} ( \mathcal{S}(\mathbf{X}))}, & \text{Mandarin}\\
    \end{cases}       
\end{equation}
where $\mathcal{S}(\cdot)$ indicates the shared language-independent module.
$\mathcal{L}_{en}(\cdot)$ and $\mathcal{L}_{cn}(\cdot)$ denote the English and Mandarin language-specific modules, respectively. 

During training, the language ID serves as a switch to select the desired language-specific module.
The whole network is trained jointly to minimize the mean square errors between the original acoustic features and the predicted ones in each language.
In each training step, only the relevant gradients will be computed to update the associated weights, while the loss in the other language will be set to zero.

\vspace{-0.2cm}
\subsection{Mixed-lingual PPG}
\label{ssec:mix}
Mixed-lingual PPG is obtained by a unified English-Mandarin acoustic model acoustic model for code-switch speech recognition.
Specifically, a time-delay neural network-based acoustic model~\cite{waibel1995phoneme,peddinti2015time} is trained with the senones as targets. 
Being exposed to speech data from both languages during training, this model learns to capture the acoustic similarity while differentiate the specific phones in the two languages.
During testing, joint posterior probability distributions are obtained by passing the test frames through the trained model.
By doing so, this unified acoustic model takes the union of two phonetic systems into one as if it were for the same phonetic system.
The dash box in Fig.~\ref{fig:network} shows the phone recognizer for mixed-lingual PPG extraction.
Instead of using two phone recognizers in each language as in Fig.~\ref{fig:vc_system}, now it only consists of a single unified English-Mandarin acoustic model.

\vspace{-0.2cm}
\section{Experiments}
\label{sec:exp}
\subsection{Database and Feature Extraction}
\label{ssec:data}
We conducted cross-lingual VC experiments between English and Mandarin languages.
10 English speakers (5 female and 5 male) from VCC2016 database~\cite{toda2016voice} and 10 Mandarin speakers (5 female and 5 male) from the Mandarin average model database\footnote{\url{http://www.data-baker.com/hc_pm_en.html}} were chosen for training.
Each speaker provided 162 utterances, where 150 utterances were used for training, while the other 12 utterances were used for validation.
In total, we had 3,000 utterances from 20 speakers for the average model training.
For testing, we selected 4 bilingual speakers (2 female and 2 male) from the EMIME Mandarin Bilingual Database~\cite{wester2011emime}.
We converted 20 utterances for each conversion pair.
Both intra-gender and inter-gender conversions were performed between the selected speakers, as summarized in Table 1.

\begin{table}
        \small
        \caption{The composition of the training and test data.}
        \centering
        \begin{tabular}{ c|c|c  }
        \hline
          & \textbf{Database} & \textbf{Selected Speakers} \\ 
         \hline
         \multirow{4}{*}{Training}
         & VCC2016~\cite{toda2016voice}   & SF1, SF2, SF3, TF1, TF2 \\
         &   \textit{English}   & SM1,SM2,TM1,TM2,TM3   
         \\\cline{2-3}
         & Average Model  & 01F, 02F, 03F, 04F, 05F\\
         &    \textit{Mandarin}   & 07M, 08M, 09M, 12M, 13M\\
         \hline
         \multirow{2}{*}{Test}
           & EMIME~\cite{wester2011emime} & MF4, MF5 \\
           & \textit{English/Mandarin} & MM1,MM2\\
         \hline
        \end{tabular} 
        \label{table:data}
\end{table}
%

All speech signals were sampled at 16kHz with a frame shift of 5ms.
The speech data were first analyzed by the WORLD vocoder~\cite{morise2016world} for feature extraction, generating the spectrum (513-dim), AP (1-dim), and $F0$ (1-dim) features.
Then the Speech Signal Processing Toolkit\footnote{\url{https://sourceforge.net/projects/sp-tk/}} was used to compute the 40-dimensional MCCs. 

Kaldi toolkit \cite{povey2011kaldi} was used to train the English, Mandarin and the unified English-Mandarin acoustic models.
A time-delay neural network with 10 hidden layers and 1280 hidden units per layer was trained with lattice-free maximum mutual information (LF-MMI)~\cite{povey2016purely} criterion using 40-dimensional mel frequency cepstral coefficients (MFCC) features of approximately 200 hours speech data from each language. 
The hyperparameters were chosen according to the standard recipe provided for the Switchboard database in the Kaldi toolkit~\cite{povey2011kaldi}. 
The acoustic data was a random subset of the open-source Librispeech~\cite{panayotov2015librispeech} and AI-SHELL2~\cite{du2018aishell} corpora.
The frame rate was adjusted to VC task and set to 5ms. 
For a similar reason, the frame subsampling rate was set to 1. 

The output layers for English and Mandarin had 4,193 and 6,360 context-dependent phone states respectively.
To compensate for the imbalance in the size of English and Mandarin phonetic alphabets due to the tonality, we have used 213 separate position-dependent models for English phones and 199 position-independent models for Mandarin phones. 
With extra silence and spoken noise entries, the English, Mandarin and unified English-Mandarin acoustic model had 215 (213 + 2), 201 (199 + 2) and 414 (213 + 199 + 2) classes, respectively.
As a result, a linguistic feature frame of either bilingual PPG or mixed lingual PPG had 414 elements.

For i-vector extractor, the universal background model (UBM) contained 502 speakers (251 female and 251 male) from Switchboard II corpus.
In total, we used 1,872 utterances, each was 5 minutes long on average.
MFCC features were used to derive the i-vectors. 
We used gender-independent UBM of 1024 mixture components and total variability matrix with 400 speaker factors. 
The i-vector dimension were reduced from 400 to 150 after applying LDA.

By combining the PPG frame of 414 elements, and an i-vector of 150 elements, we obtain an input feature frame of 564 dimensions.
The output acoustic feature contained 127 elements including a voiced/unvoiced flag (1-dim), the original and their delta and delta-delta coefficients of MCCs (40-dim), log $F0$ (1-dim), AP (1-dim).

\subsection{Experimental Setup}
\label{ssec:setup}
We have implemented 4 different cross-lingual VC systems for comparison. The details of the baseline and proposed systems were discussed as follows.
\begin{itemize}[leftmargin=*]
\itemsep0em
    \item \textbf{bPPG-LI:}
    The cross-lingual VC system~\cite{zhou2019cross} using bilingual PPG (bPPG) with a single language-independent (LI) output layer as discussed in Section~\ref{ssec:system}. The model consisted of two BLSTM layers with 256 hidden units in each layer.
    This is benchmarked as our baseline.
    \item \textbf{mPPG-LI:} 
    The cross-lingual VC system using mixed-lingual PPG (mPPG) introduced in Section~\ref{ssec:mix} with a single language-independent (LI) output layer. 
    The model used the same network configurations as bPPG-LI.
    \item \textbf{bPPG-LS:}
    The cross-lingual VC system using bilingual PPG (bPPG) and the proposed modularized neural network with language-specific (LS) output layers as discussed in Section~\ref{ssec:network}.
    All shared layers had 256 nodes including one input ReLU layer and two hidden BLSTM layers.
    Both language-specific output ReLU layers had 128 nodes. 
    \item \textbf{mPPG-LS:}
    The cross-lingual VC system using mixed-lingual PPG (mPPG) based on our proposed modularized neural network with language-specific (LS) output layers. 
    The model used the same hyper-parameters as bPPG-LS.
\end{itemize}

Merlin toolkit~\cite{wu2016merlin} was used for model training. 
All models were trained with a learning rate of 0.002, the minibatch size and momentum were set as 25 and 0.9 respectively.

During conversion, $F0$ were converted by a global linear transformation in log-scale~\cite{berrakis2018}. 
APs were directly copied from source speech, and the MCCs were generated by Maximum Likelihood Parameter Generation algorithm~\cite{tokuda2000speech}. 
The post-filtering processing technique in the cepstral domain was used to enhance the conversion quality~\cite{takayo2005post}.

\subsection{Objective Evaluation}
\label{ssec:obj}
For objective evaluation, mel-cepstrum distortion (MCD) was used to measure the spectral distance between the converted and original speech from bilingual speakers, defined as
%
\begin{equation} \label{eq:3}
MCD[dB] = 10/\mathrm{log}10\sqrt{2\sum_{d=1}^{D}(\hat{Y}_d-Y_d)^2},
\end{equation}
where $D$ is the MCC feature dimension, $\hat{Y}_d$ and $Y_d$ are the $d^{\mathrm{th}}$ coefficients of the converted and original MCCs, respectively.
A lower value accounts for a smaller distortion.
\begin{figure}
	\centerline{\includegraphics[width=.5\textwidth]{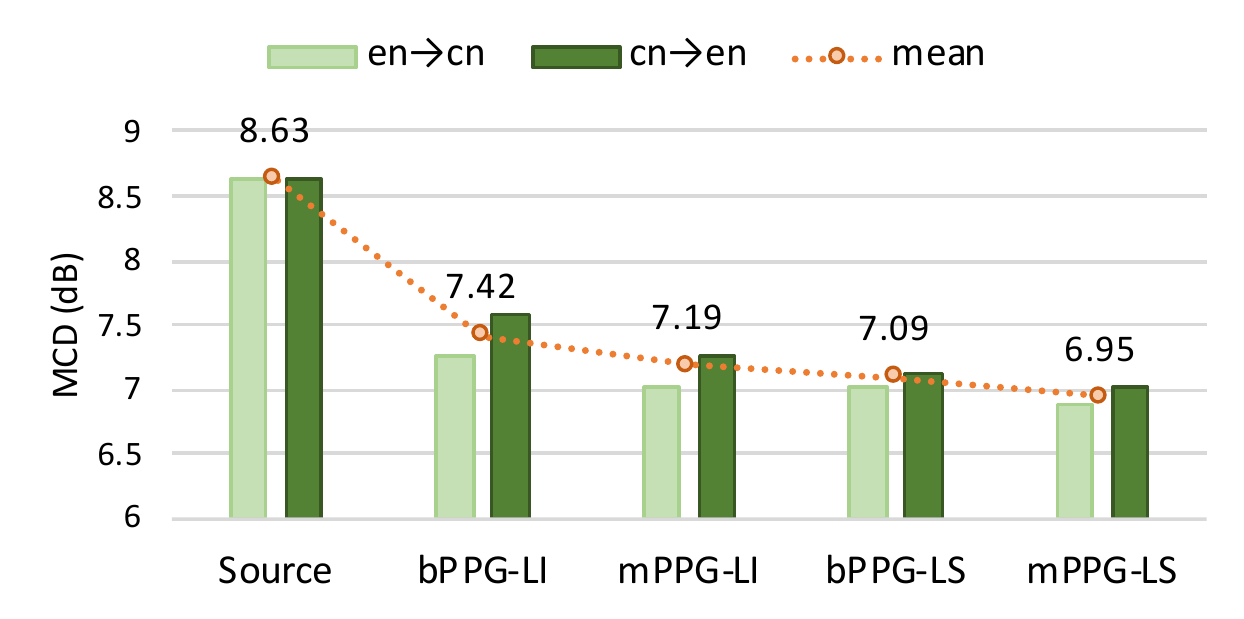}}
	\vspace{-0.2cm}
	\caption{Average MCD results for cross-lingual VC using bilingual test speakers. 
	Source indicates the source speech without conversion.en $\rightarrow$ cn means that we convert a source English utterance to a Mandarin target speaker, and vice-versa.
	}
	\label{fig:MCD}
\end{figure}

Average MCDs of cross-lingual VC are presented in Fig.~\ref{fig:MCD}.
First, we compare the proposed modularized neural network using the language-specific output layers with the baselines. 
It is observed that both bPPG-LS and mPPG-LS outperform their counterparts (bPPG-LI and mPPG-LS) with the average MCD dropping from 7.42dB to 7.09dB and 7.19dB to 6.95dB, respectively.
It suggests that the proposed modularized neural network yields better conversion performance than the baselines, in terms of MCD.
Then we further evaluate the performance of the proposed mixed-lingual PPG.
It can be found that systems using mixed-lingual PPG (mPPG-LI and mPPG-LS) consistently outperform those using bilingual PPG (bPPG-LI and bPPG-LS).

\begin{figure}
	\centerline{\includegraphics[width=.5\textwidth]{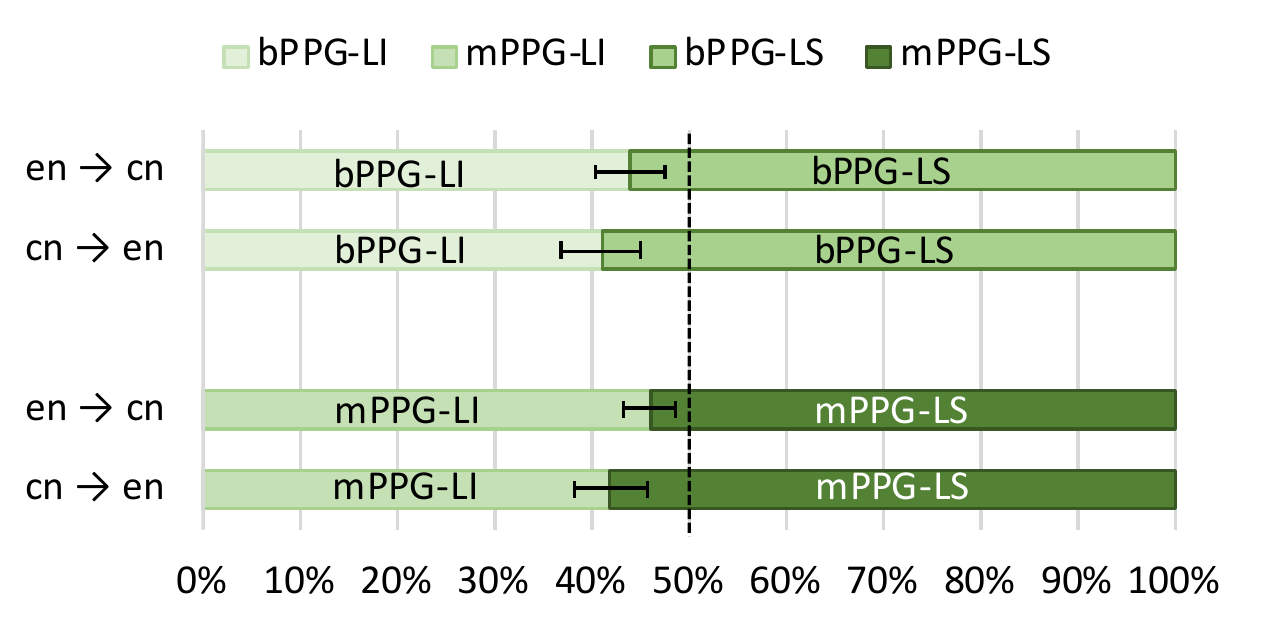}}
	\vspace{-0.2cm}
	\caption{AB preference test results for speech quality with $95\%$ confidence intervals between systems using proposed language-specific and the baseline language-independent output layers (i.e., bPPG-LI vs. bPPG-LS; and mPPG-LI vs. mPPG-LS). en $\rightarrow$ cn means that we convert a source English utterance to a Mandarin target speaker, and vice-versa.}
	\label{fig:AB_network}
\end{figure}

\subsection{Subjective Evaluation}
\label{ssec:sub}
We further conducted listening tests.
AB preference and mean opinion score (MOS) tests were conducted to assess speech quality.
Meanwhile, VCC similarity tests, the Same/Different paradigm from the VCC 2016~\cite{toda2016voice}, were conducted to assess speaker similarity.
In addition, to further study listeners' preferences across all VC systems, we also conducted best-worst scaling (BWS) tests~\cite{sisman2019group} on both speech quality and speaker similarity.
12 samples were randomly selected from (4 x 20 = 80) converted samples of each VC system. 
12 English and Mandarin bilingual listeners participated all tests.

\begin{itemize}[leftmargin=*]
\item[a)] \emph{AB preference tests:} In AB preference tests, A and B were the randomly selected converted speech samples from different systems, and the listeners were asked to compare their voice quality and naturalness.

First, we examine the effectiveness of the proposed modularized neural network. 
The speech quality test results are presented in Fig.~\ref{fig:AB_network}.
It is observed that mPPG-LS and bPPG-LS significantly outperform mPPG-LI and bPPG-LI, respectively.
And the performance improvement is more remarkable in en $\rightarrow$ cn than that in cn $\rightarrow$ en.
The test results demonstrate that our proposed modularized neural network is more effective than the baseline network using a single output layer in terms of speech quality.
\begin{figure}
	\centerline{\includegraphics[width=0.5\textwidth]{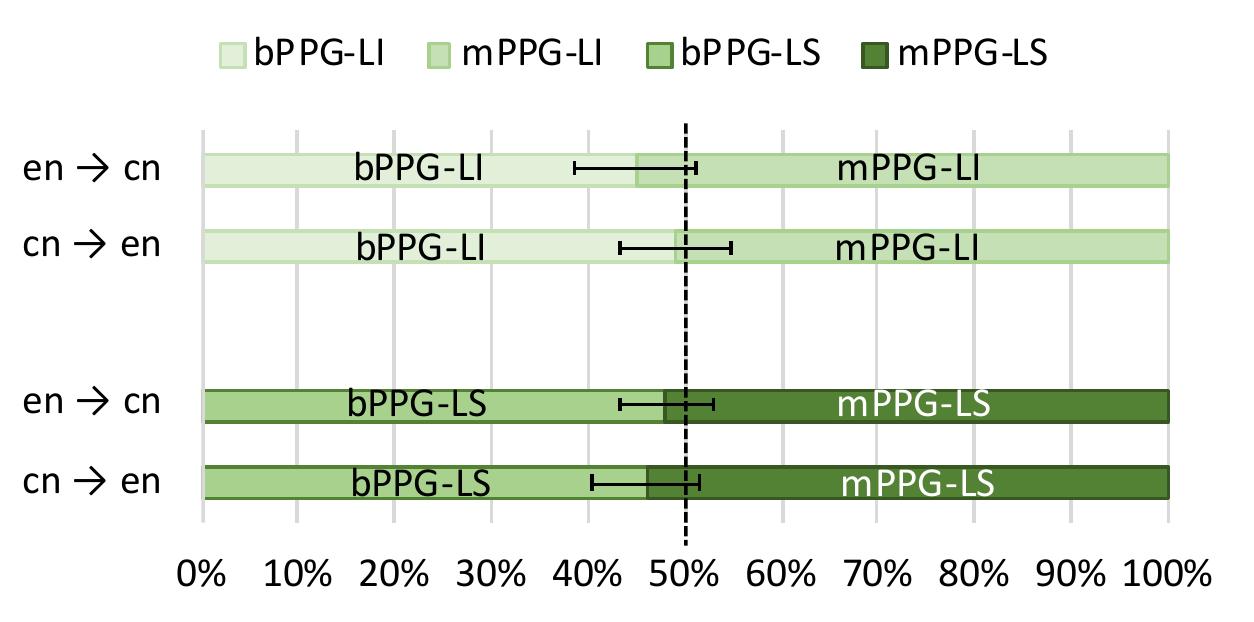}}
	\vspace{-0.4cm}
	\caption{AB preference test results for speech quality with $95\%$ confidence intervals between systems using proposed mixed-lingual PPG and baseline bilinugal PPG (i.e., bPPG-LI vs. mPPG-LI; and bPPG-LS vs. mPPG-LS).
	en $\rightarrow$ cn means that we convert a source English utterance to a Mandarin target speaker, and vice-versa.}
	\label{fig:AB_ppg}
\end{figure}

Then, we further compare the proposed mixed-lingual PPG with bilingual PPG.
Fig.~\ref{fig:AB_ppg} shows the speech quality test results, which suggests that systems using proposed mixed-lingual PPG outperform those using bilingual PPG in both en $\rightarrow$ cn and cn $\rightarrow$ en, though the difference is not statistically significant.
The speech quality test results confirm that we can consider mixed-lingual PPG as a better representation than bilingual PPG to characterize the linguistic information from two languages.

\begin{table}
        \caption{MOS test results over all cross-lingual VC systems.
        en $\rightarrow$ cn means that we convert a source English utterance to a Mandarin target speaker, and vice-versa. 
        }
        \centering
        \begin{tabular}{ c | c | c  }
        \hline
          \textbf{System} & \textbf{en $\rightarrow$ cn} & \textbf{cn $\rightarrow$ en}  \\ 
         \hline
         bPPG-LI & 2.87 $\pm$ 0.16 & 2.42 $\pm$ 0.72 \\
         mPPG-LI & 2.98 $\pm$ 0.24 & 2.45 $\pm$ 0.58 \\
         bPPG-LS & 3.55 $\pm$ 0.28 & 3.43 $\pm$ 0.38 \\         
         \textbf{mPPG-LS} & \textbf{3.62} $\pm$ 0.28 & \textbf{3.53} $\pm$ 0.37 \\
         \hline
        \end{tabular} 
        \label{table:mos}
\end{table}

\item[b)] \emph{MOS tests:} In MOS tests, listeners were asked to rate the quality and naturalness of the converted speech on a 5-point scale. 
The results are presented in Table.~\ref{table:mos}.
It can be firstly observed that mPPG-LS obtained the highest scores in both en $\rightarrow$ cn and cn $\rightarrow$ en.
We also find that bPPG-LS and mPPG-LS achieve significant improvement over bPPG-LI and mPPG-LI, respectively.
The results suggest that our proposed modularized neural network using mixed-lingual PPG is fairly effective in cross-lingual VC.

\begin{figure}[t]
	\centerline{\includegraphics[width=0.5\textwidth]{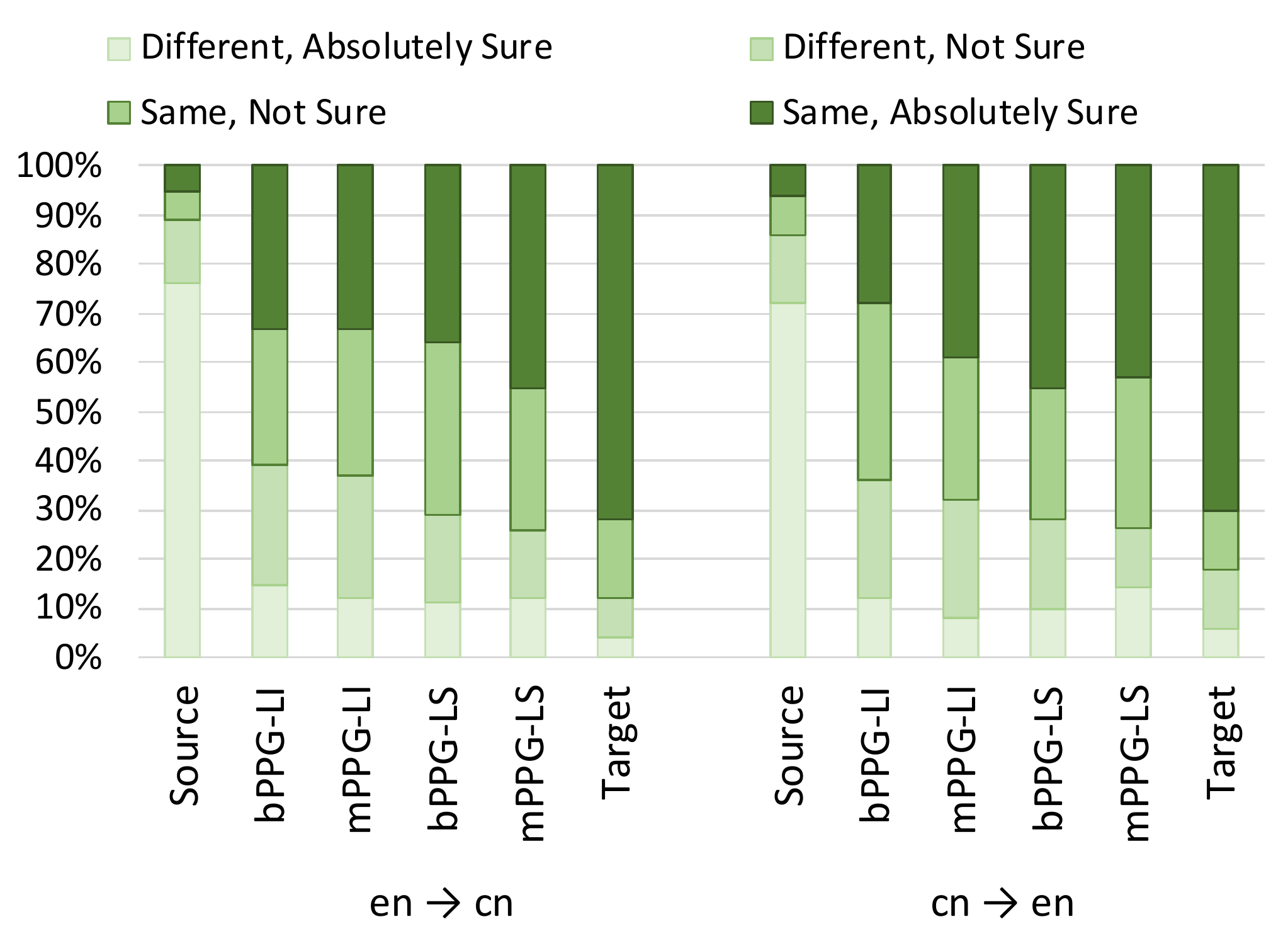}}
	\vspace{-0.5cm}
	\caption{VCC similarity test result, en $\rightarrow$ cn means that we convert a source English utterance to a Mandarin target speaker, and vice-versa. Source and target denote the source speaker and target speaker respectively.}
	\label{fig:VCC}
\end{figure}
\item[c)] \emph{VCC similarity tests:} In VCC similarity tests, the listeners were asked to compare and select whether the converted samples were uttered by the same target speaker~\cite{toda2007voice}. 
Fig.~\ref{fig:VCC} presents the similarity test results.
The system performances, ranked in ascending order (from the lowest to the highest), are as follows: bPPG-LI, mPPG-LI, bPPG-LS, and mPPG-LS.
The results are consistent in both en $\rightarrow$ cn and cn $\rightarrow$ en conversions.
It therefore demonstrates the capability of our proposed approaches for converting the speaker's identity in cross-lingual VC tasks.

%
\begin{table}
        \small
        \caption{BWS test results on 
        (a) speech quality; (b) speaker similarity.
        N/P means no preference.}
        \centering
        \begin{tabular}{ c | c | c | c | c }
        \hline
         \textbf{Test} & \textbf{System} & \textbf{Best (\%)} & \textbf{Worst (\%)} &\textbf{N/P (\%)} \\ 
         \hline
         \multirow{4}{*}{(a)}
         & bPPG-LI & 4 & 42 & 54 \\
         & mPPG-LI & 14 & 20 & 65 \\         
         & bPPG-LS & 27 & 22 & 51 \\
         & \textbf{mPPG-LS} & \textbf{55} & \textbf{16} & 29 \\
         \hline
         \multirow{4}{*}{(b)}
         & bPPG-LI & 16 & 44 & 40 \\
         & mPPG-LI & 19 & 25 & 56 \\ 
         & bPPG-LS & 22 & 20 & 58 \\
         & \textbf{mPPG-LS} & \textbf{43} & \textbf{11} & 46 \\
         \hline
        \end{tabular} 
        \label{table:bws}
\end{table}
\item[d)] \emph{BWS tests:} In BWS tests, we presented 4 converted samples with the same content from each VC system to the listeners.
In the quality tests, we asked listeners to compare all the samples and select the best and worst samples in terms of speech quality.
In the similarity tests, we provided the target speech as a reference.
The listeners were asked to select the most and the least similar samples comparing to the reference in terms of speaker identity. The BWS test results are presented in Table~\ref{table:bws}. 
In both quality and similarity tests, mPPG-LS takes the highest percentages (55\% and 43\%) as the best-converted sample among all VC systems.
While, it is chosen as the worst sample at the lowest percentages (16\% and 11\%).
The results also indicate that bPPG-LI performs the worst while bPPG-LS and mPPG-LI achieve the moderate results.

According to the subjective tests, the proposed modularized neural network using mixed-lingual PPG (mPPG-LS) consistently outperformed all the rest VC systems in both speech quality and speaker similarity tests, which demonstrates the effectiveness of our proposed approaches for cross-lingual voice conversions.
The converted samples can be found from this demo link\footnote{ {\url{https://vcsamples.github.io/asru_2019/}}}.
\end{itemize}

\section{Conclusion}
\label{sec:conclusion}
In this paper, we proposed a cross-lingual voice conversion framework based on a modularized neural network using mix-language PPG.
By utilizing the shared input module to be language independent while decomposing the output modules to be language specific, the network is robust for output acoustic feature modeling across different languages. Meanwhile, the mixed-lingual PPG extracted from a unified English-Mandarin acoustic model also provides an accurate linguistic representation for conversion quality enhancement. Experimental results successfully demonstrate our proposed approaches outperform the baseline approaches on both speech quality and speaker similarity. 

\vspace{-0.2cm}
\section{Acknowledgement}
\label{sec:ack}
This research is supported by the Agency for Science, Technology and Research (A*STAR) under its AME Programmatic Funding Scheme (Project No. A18A2b0046), the National Research Foundation Singapore under its AI Singapore Programme (Award Number: AISG-100E-2018-006), and the NUS Start-up Grant FY2016 Non-parametric Approach to Voice Morphing. Yi Zhou is also funded by the NUS research scholarship.

\small

\bibliographystyle{IEEEbib}
\bibliography{strings,refs}

\end{document}